\documentclass[aps,pra,twocolumn,showpacs,superscriptaddress,preprintnumbers,amsmath,amssymb]{revtex4}
\usepackage{graphics}
\usepackage{graphicx}
\usepackage{dcolumn}
\usepackage{bm}
\usepackage{amsmath}

\newcommand{\bra}[1]{\langle #1|}
\newcommand{\ket}[1]{|#1\rangle}
\newcommand{\braket}[2]{\langle #1|#2\rangle}
\renewcommand{\v}[1]{\mathbf{#1}}

\begin{document}

\title{A quasi-Hermitian pseudopotential for higher partial wave scattering}

\author{Iris Reichenbach}
\email{irappert@unm.edu}
\affiliation{{Department of Physics and Astronomy, University of New Mexico, Albuquerque, New Mexico 87131, USA}}
\author{Andrew Silberfarb}
\email{drews@unm.edu}
\affiliation{{Department of Physics and Astronomy, University of New Mexico, Albuquerque, New Mexico 87131, USA}}
\author{Ren\'e Stock}
\email{rei@qis.ucalgary.ca}
\affiliation{{Institute for Quantum Information Science, University of Calgary, Alberta T2N 1N4, Canada}}
\author{Ivan H. Deutsch}
\email{ideutsch@unm.edu}
\affiliation{{Department of Physics and Astronomy, University of New Mexico, Albuquerque, New Mexico 87131, USA}}



\begin{abstract}
We formulate a new quasi-Hermitian delta-shell pseudopotential for higher partial wave scattering, and show that any such potential must have an energy-dependent regularization. The quasi-Hermiticity of the Hamiltonian leads to a complete set of biorthogonal wave functions that can be used as a basis to expand and diagonalize other two-body Hamiltonians. We demonstrate this procedure for the case of ultracold atoms in a polarization-gradient optical lattice, interacting pairwise when two atoms are transported together from separated lattice sites. Here the pseudopotential depends explicitly on the trapping potential. Additionally, we calculate the location of trap-induced resonances for higher partial waves, which occur when a molecular eigenstate is shifted to resonance with a trap eigenstate. We verify the accuracy of the pseudopotential approach using a toy model in which a square well acts as the true interaction potential, and see excellent agreement.
\end{abstract}

\maketitle

\section{Introduction}
The behavior of trapped ultracold neutral atoms has attracted considerable interest in both quantum condensed matter physics and quantum information science.  In these disciplines, ultracold atoms in tightly confining traps are an attractive platform given the richness of the systems and the ability to coherently control a wide variety of parameters in a precise and well understood way. From a condensed matter physics perspective, ultracold atoms trapped in optical lattices provide an ideal system for studying tight-binding models such as the Hubbard model \cite{deutsch1998,jaksch1998,hofstetter2002}.  In the context of quantum information, optical trap-arrays provide a scalable platform for storing many qubits, with parallel operations, applicable to quantum-cellular automata \cite{brennen2003,raussendorf2005}, or more general quantum circuit operations \cite{brennen1999}, and fault-tolerance via topological encoding \cite{raussendorf2006}.

At the fundamental level, quantum many-body systems are governed by their two-body interactions.  In the case of ultracold neutral atoms, these consist of collisions determined by the diatomic molecular interaction potential. Coherences between unbound scattering atoms and bound molecules provide new handles for tuning interactions and exploring heretofore uncontrollable many-body phenomenen such as superfluid to Mott insulator transitions \cite{greiner02} and the BCS to BEC crossover \cite{koetsier2006}.  The production of ultracold molecules also opens the door to new studies in quantum chemistry where nuclear motion is fundamentally quantum and controllable.  Additional novel phenomena arise in tightly confining traps due to reduced dimensionality including the Tonks Gas for strongly interacting 1D Bose systems \cite{paredes2004} and trap-induced resonances \cite{stock2003,stock2006}.     

For ultracold collisions, the Wigner-threshold law dictates that the lowest nonvanishing partial wave dominates the interaction.  Unless ruled out by exchange symmetry, $s$-wave collisions dominate and are generally characterized by one parameter, the scattering length $a$.  This can be codified into a contact ``pseudopotential", whose scattering phase shift in the Wigner-threshold regime equals that of the true molecular potential.  The use of the Fermi pseudopotential well-describes the properties of degenerate quantum gases, such as the mean-field effects predicted by the Gross-Pitaevskii equation \cite{blume2002}.   The case of two atoms in a trap can be treated with great efficacy by the same pseudopotential as for atoms scattering in free space.  The contact potential effectively sets the boundary condition for the relative coordinate wave function at short range \cite{busch,bolda2002,blume2002} whereas the trap sets the long-range boundary conditions.  The resulting two-atom problem determines a bound-state energy spectrum and corresponding set of eigenfunctions that can serve as a basis for the studies of many-body systems with pairwise interactions in a trap {\cite{stock2003, stock2006, bolda2002, bolda2003, blume2002}.  

Although $s$-wave collisions typically dominate at ultracold temperatures, in some cases higher partial waves play an important role in the interaction. For example, in the case of identical fermions, $s$-wave scattering is forbidden and $p$-wave scattering dominates. For the boson $^{133}$Cs there are anomalous collision properties due to bound state of the dimer near zero energy leading to large spin relaxation through coupling to higher partial waves \cite{leo2000}.
The generalization to higher partial wave scattering has recently been revisited by numerous authors. A long-standing error in the Huang pseudopotential for arbitrary partial waves \cite{huang} has been corrected \cite{stock2005, roth2001, kanjilal2004,idziaszek2006, pricoupenko2006}.
Our solution \cite{stock2005}, based on a delta-shell potential at finite range, allows for arbitrary multipole distributions associated with higher-order partial waves. This has been used to calculate the level shifts of the eigenstates of trapped atoms in the context of optical lattices \cite{stock2003, stockdissertation}. 

For $s$-waves, the pseudopotential also gives rise to normalizable wave functions that can be used to diagonalize other Hamiltonians describing interactions between trapped atoms, e.g. two atoms in separated traps of a polarization-gradient lattice \cite{stock2003,stockdissertation}. Such a system exhibits new trap-induced resonances, which occur when a molecular bound state of the interaction potential is shifted to the same energy as a vibrational state of the trap \cite{stock2003}. These trap-induced resonances offer a new handle for the control of interactions between ultracold atoms, and for the design of quantum logic gates.  In other examples, generalized pseudopotentials have been used to study higher-order partial wave interactions in anisotropic traps, including the approach to quasi-2D gases \cite{idziaszek2005,kanjilal2004}. 

In this article, we introduce a new quasi-Hermitian higher-partial wave pseudopotential that allows us to construct a complete biorthonormal set of eigenfunctions. This is essential for the application of the pseudopotential to trapped atoms, and in particular to the correct description of higher partial wave trap-induced resonances. Whereas our previously derived pseudopotential~\cite{stock2005} based on a delta shell allows a correct derivation of the eigenspectrum of interacting atoms, as reviewed in Sec.\ \ref{sec:review}, the eigenfunctions for this non-Hermitian pseudopotential do not form an orthonormal set. We show that even in the Wigner regime with constant higher partial wave scattering lengths, we need an energy-dependent regularization to correct this problem. In Sec. \ref{sec:non-Hermitian} we construct the appropriate quasi-Hermitian potential. The time-independent Schr\"{o}dinger equation is then solved self-consistently, giving rise to a biorthonormal set of wave functions, as discussed in Sec. \ref{sec:doublyselfconsistent}.  We apply our solution to the study of trap-induced resonances in polarization-gradient optical lattices, generalized to higher-partial wave scattering in Sec.\ref{sec:tisr}.  Our results are tested using a toy model, where a square well acts as the true interaction potential, and find excellent agreement. Section \ref{sec:summary} summarizes our results.

\section{Review of the Pseudopotential Construction}
\label{sec:review}
Ultracold atoms can have a very long deBroglie wave length, on the order of microns, whereas the van der Waals interaction for alkali atoms is on the order of 1-10 \AA \cite{insightcold}. For scattering from a short-range, radially symmetric potential, the effect of the interaction is to phase shift the radial wave function so that it takes the asymptotic form
\begin{equation}
F^{>}_{l,k}(r)=A_l \left(j_l (kr) - \tan{(\delta_l(k))} n_l (kr) \right) , 
\label{asymptotic}
\end{equation}
where $\delta_l$ is the scattering phase shift, $k$ is the wave number associated with the relative coordinate kinetic energy, and $j_l$, $n_l$, are the spherical Bessel and Neumann functions, respectively.  It is possible to replace the true interaction potential with a simpler potential, as long as it gives the same asymptotic wave function. For $s$-waves in the Wigner threshold regime, where the scattering phase shift is related to the scattering length by $\tan  \delta_0 \approx -ka$, this can be done using a regularized Fermi's pseudopotential, defined as
\begin{equation}
V_0 (\v{r}) = 2 \pi a \; \delta^{(3)} (\v{r}) \frac{\partial}{\partial r }r.
\label{swavepseudopotential}
\end{equation}
Here and throughout, we take units with $\hbar =\mu=1$, where $\mu$ is the reduced mass of the two-atom system. The first order derivative serves to regularize the pseudopotential by removing the singular contributions at short range, allowing it to act on arbitrary wave functions, including those that diverge at the origin,  e.g., the Neumann functions $n_l$ in Eq.\ (\ref{asymptotic}). The only parameter needed for reproducing the actual asymptotic wave function is the zero-energy scattering length, $a$, given by $a= -\lim_{k\rightarrow 0}\left[ \tan{(\delta_0)}/k \right]$.

For two atoms in a tightly confining trap, the interaction causes a shift in the unperturbed vibrational states.  There are three length scales that govern the physics: the range of the interaction potential, $\beta_6= (m C_6/\hbar^2)^{1/4}$  for van der Waals interactions, the scattering length, $a$, and the width of the unperturbed ground vibrational state in the trap, $r_0$. These length scales are often very different, making it difficult to find a numerical solution to the Schr\"{o}dinger equation for geometries that are not separable into 1D solutions. A solution can be found by modeling the net effect of the interaction by the same pseudopotential as in free space if the confining trap does not significantly distort the short-range interaction potential, true when $\beta_6 \ll r_0$, which is usually the case. This approach has the advantage of separating the short range behavior from  the long range solutions. 

In a seminal paper, Busch {\em et al.} \cite{busch} solved for the new energy eigenfunctions and values for two atoms in a harmonic trap interacting via $s$-wave collisions, modeled using the regularized Fermi pseudopotential, parameterized by the free-space scattering length $a$.  The solution is found in the transcendental equation, $a = \Gamma(-\nu -1/2) / 2\Gamma{(-\nu)}$ with $E = 2 \nu +3/2$, having chosen units with the trap frequency $\omega=1$ so that all lengths are measured in $r_0=\sqrt{\hbar/\mu \omega}$ and energy in units of $\hbar \omega$.  As long as the vibrational energy is within the Wigner threshold regime, guaranteed when $a\ll r_0$, this accurately captures the perturbed spectrum.  Out of that regime, one employs a free-space pseudopotential, parameterized by an {\em energy-dependent} scattering length, $a \rightarrow a(E)=-\tan\delta_0(k)/k$, where $E=k^2/2$. Since now the strength of the pseudopotential depends on the vibrational energy, whose value we seek, the Schr\"{o}dinger equation must be solved self consistently \cite{bolda2002,blume2002}. One solves for the energy as a function of scattering length $E(a)$ (the ``Busch solutions") and separately $a(E)$ in a {\em free space} scattering calculation. This leads to the following set of equations, 
\begin{subequations}
\begin{align}
 -\frac{\tan{(\delta _0(k))}}{k} &=\frac{1}{2} \frac{\Gamma(-\nu -1/2)}{\Gamma (-\nu)}, \\
\frac{1}{2}k^2 &= 2 \nu + 3/2.
\end{align}
\end{subequations}
These solutions show excellent agreement with full scattering calculations \cite{bolda2002}.

For higher partial waves, this construction is more subtle. The potential is separable into a radial and an angular part, with
\begin{equation}
V(\vec r) = v_l(r) Y_{l m}(\hat r),
\end{equation}
where $Y_{l m}$ are the spherical harmonics.
A contact potential at the origin cannot support higher-order multipoles, $l>0$.  Thus, no pseudopotential for higher partial waves can be constructed exactly at the origin.  Derevianko \cite{derevianko2005} solved this by employing the appropriate Green's identity in order to relate a surface integral at a finite radius to the pseudopotential.  Our solution explicitly defines the pseudopotential as the limiting form of a finite radius $\delta$-shell \cite{stock2005}.  Defining higher-order scattering lengths by the Wigner threshold law,
 $\tan{\delta_l (k)} \approx -(k a_l)^{2l+1}$, the $\delta$-shell pseudopotential takes the form,
\begin{equation}
V_l (r)= \lim_{r_s\rightarrow0}  a_l^{2l+1} \frac{\delta(r-r_s)}{r_s^{l+2}} \hat{D_l},
\label{renespotential}
\end{equation}
where 
\begin{equation}
\hat{D_l} = \frac{(2l+1)!!}{2 (2l)!!}\frac{1}{r^{l+2}}\frac{\partial^{2l+1}}{\partial r^{2l+1}}r^{l+1}
\label{regularizationoperator}
\end{equation}
is the regularization operator.  For $s$-waves, one recovers the familiar regularized Fermi pseudopotential. 

As demonstrated in~\cite{stock2005}, the generalization of the Busch solutions to arbitrary partial-wave scattering follows from Eq.\ (\ref{renespotential}), parameterized by the energy-dependent $l$-wave scattering phase shift in free space according to the substitution $a_l^{2l+1} \rightarrow -\tan \delta_l(k)/k^{2l+1} \equiv \beta_l(k)$ (see also \cite{peach2004} for a quantum-defect theory approach). An ansatz for the radial wave function of the form $F(r)= r^{l} e^{-r^{2}/2}w(r)$ transforms the piecewise quadratic radial equation into Kummer's equation, with confluent hypergeometric functions $M$ and $U$ as solutions \cite{abramowitz, stockdissertation}.  Inside the shell, the solution must be the Kummer function that is regular at the origin, while outside, the boundary equation at infinity requires the solution which asymptotically approaches zero,
\begin{subequations}
\begin{align}
w^{<}(r) &= B_l\; M(-\nu,l+3/2,r^2) \, & r<r_s, \\
w^{>}(r) &=  A_l\; U(-\nu,l+3/2,r^2) \, & r>r_s, 
\end{align}
\label{Kummer}
\end{subequations}
where the energy eigenvalue is $E=2 \nu + l +3/2$ in harmonic oscillator units.  Note, when $\nu$ is an integer, the solution everywhere is the familiar Laguerre polynomial for an unperturbed isotropic harmonic oscillator; noninteger $\nu$ arise from the pseudopential, which shifts the energies of the trap eigenfunctions. The boundary conditions for the wave function across the delta-shell,
\begin{align}
F^{<}(r_s) & =  F^{>}(r_s) , \nonumber \\
\frac{1}{2} \left( \frac{d F^{<}}{dr} -\frac{d F^{>}}{dr}  \right)_{r_s}  & =   a_l^{2l+1} \hat{D_l}F^{>}(r_s)  , 
\label{boundaries_for_Busch}
\end{align}
determine the wave functions and a transcendental equation for the eigenvalue $\nu$,
\begin{multline}
\beta_l(k)= -\frac{\tan \delta_l(k)}{k^{2l+1}}= \\
(-1)^l \frac{2}{\pi} \left[ \frac{\Gamma(l+3/2)}{(2l+1)!!} \right]^2 \frac{\Gamma(-\nu-l-1/2)}{\Gamma(-\nu)}.
\label{general_Busch}
\end{multline}
The generalized, self-consistent Busch eigenvalues are those for which $E=2\nu + l + 3/2 = k^2/2$.  

\section{A quasi-Hermitian pseudopotential}
\label{sec:non-Hermitian}
Whereas the higher partial-wave pseudopotential suggested in \cite{stock2005} allowed us to derive the generalized eigenvalue spectrum for trapped atoms, it fails, as we will show in the following, in the construction of a complete set of orthonormal eigenfunctions.
This task is stymied by the regularization operator of the pseudopotential Eq.\ (\ref{regularizationoperator}) which contains odd powers of derivative operators, rendering it non-Hermitian for all but $s$-waves \footnote{The $s$-wave pseudopotential is only Hermitian due to the cancellation of its anti-Hermitian part by the kinetic energy part of the Hamiltonian as discussed in the Appendix}. Its eigenfunctions do not form an orthonormal set, as can be seen by considering the limit of the eigenfunctions when the shell radius $r_s$ of the potential is taken to zero. According to Eq.\ (\ref{renespotential}), the  pseudopotential for $p$-waves has the form
\begin{equation}
\lim_{r_s\rightarrow0} C \frac{1}{r^3}\delta(r-r_s)\frac{\partial^3}{\partial r^3}r^2,
\end{equation}
where $C$ is a constant. For $l>0$, the reduced radial wave functions outside the shell, $r F^{>}(r)$ in Eq.\ (\ref{Kummer}), diverge as $r_s^{-l}$, for $r_s \rightarrow 0$. Therefore, most of the probability is centered around $r=r_s$. Normalizing these wave functions makes the oscillatory tail negligible as $r_s\rightarrow 0$, and hence, the integral over wave functions associated with different energies cannot be zero because the pronounced peak common to all wave functions cannot be canceled. In contrast, for $s$-waves the outside reduced radial functions $rF^{>}(r)$ go to a finite value at $r\rightarrow0$, and hence, we can take the limit of zero shell radius without obtaining this pronounced peak. In fact, in the limit $r_s\rightarrow0$ the pseudopotential for $s$-waves has an {\em orthonormal}, complete set of eigenvectors with real eigenvalues and is therefore Hermitian. For a more rigorous proof, see Appendix \ref{$s$-wavehermiticity}.

We can obtain a complete set of eigenfunctions associated with a pseudopotential for higher partial waves in the following way. Since the Hamiltonian has real eigenvalues, it is possible to construct a {\em quasi-Hermitian} potential, with a {\em biorthonormal} set of eigenvectors, 
\begin{equation}
H=\sum_nE_n\ket{\Psi_n}\bra{\Phi_n},
\end{equation}
such that $H \ket{\Psi_n} = E_n \ket{\Psi_n}$, $\bra{\Phi_n} H^\dagger = \bra{\Phi_n}E_n $, and $\braket{\Phi_n}{\Psi_m}=\delta_{nm}$ \cite{geyer92}.
This, however, requires a new regularization since a delta-potential with an odd number of derivatives beyond first order does not possess a quasi-Hermitian adjoint with finite extension.  Projecting onto the partial wave $l$, we take a new ansatz for the radial dependence of the  pseudopotential restricted to a regularization with at most a first order derivative,\begin{equation}
V_l(r) = c_1 \tan\delta_l(k) \delta(r-r_s)   \left[ c_2 +\frac{\partial}{\partial r} \right] ,
\label{quasihermitianpseudopotential}
\end{equation}
where $c_1$ and $c_2$ are constants to be determined and $\delta_l (k)$ is the scattering phase shift at kinetic energy $E_k=k^2/2$.

We begin by constructing a higher partial wave pseudopotential for atoms scattering in free space, where the radial wave function corresponding to the eigenstate of the pseudopotential takes the form Eq.\ (\ref{asymptotic}) for $r>r_s$ and $F_l^{<}(r) = B_l j_l(kr)$ for $r<r_s$.   The constant $c_2$ should be chosen in such a way as to remove the irregular component of the wave function outside the shell,
\begin{equation}
\left[c_2 +\frac{\partial}{\partial r}\right]_{r_s} n_l(r)=0.
\label{regularizationcondition}
\end{equation}
Additionally, we must satisfy the boundary condition given by integration of the Schr\"{o}dinger equation over the shell radius,
\begin{multline}
\frac{1}{2}\left.\left[\left. \frac{\partial}{\partial r} F^>\right|_{r_s} - \left. \frac{\partial}{\partial r} F ^< \right|_{r_s} \right]  + c_1 c_2 \tan{(\delta_l(k))} F^>\right|_{r_s} +\\
c_1  \tan{(\delta_l (k))}  \left. \frac{\partial}{\partial r}F^> \right|_{r_s}=0.
\label{boundaryfromschroedinger}
\end{multline}
These conditions can be achieved only with a regularization that depends explicitly on the {\em kinetic energy} of the colliding particles.  As we have already considered the extension of the pseudopotential with an energy-dependent scattering length to account for situations beyond the Wigner-threshold regime, the extension to an energy-dependent regularization is natural. Though not unique, one possible choice for $c_1$ and $c_2$ is
\begin{equation}
c_1=-\frac{n_l(kr_s)}{j_l(kr_s)},  \, c_2=-\frac{n'_l(kr_s)}{n_l(kr_s)}.
\end{equation}
For small shell radii, and for $s$-wave scattering ($l=0$), $c_1 \rightarrow -1/kr$, $c_2 \rightarrow 1/r$.  The resulting pseudopotential is seen to recover the familiar Fermi pseudopotential  for $r_s \rightarrow 0$ in the Wigner threshold regime with $\tan (\delta_0(k)) \rightarrow -ka$ with $\delta (r-r_s) \rightarrow 4 \pi r^2 \delta^{(3)}(r)$.
Note that given the wave functions' dependence on the kinetic energy, the pseudopotential for higher partial waves cannot be reduced to an energy-independent form, even in the Wigner-threshold regime.  As a result we must solve the eigenvalue equation self-consistently, as will be discussed in Sec. \ref{sec:doublyselfconsistent}.

This quasi-Hermitian pseudopotential can be extended to the case of interacting atoms in traps.  The radial wave functions take the canonical form that generalizes Eq. (\ref{asymptotic}),
\begin{subequations}
\begin{align}
F_{l,E}^{<}(r) &= B_l f_{l,E}(r),  \, & r<r_s, \\
F_{l,E}^{>}(r) &=  A_l(f_{l,E}(r) +  \beta_l(k) g_{l,E}(r) ), \, & r>r_s,
\end{align}
\label{eigenfunctions}
\end{subequations}
where $f_{l,E}(r)$ and $g_{l,E}(r)$ are respectively the solutions to the Schr\"{o}dinger equation at energy $E$ that are regular and irregular at the origin, and $\beta_l(k) = -\tan(\delta_l(k))/k^{2l+1}$  is the scattering length coefficient. The irregular solutions are chosen in such a way, that as $r \rightarrow 0$, the ratio $g_{l,E}(r)/f_{l,E}(r)$ behaves as the ratio of the wave functions in free space $j_l(k r)/n_l(k r)$. The general form of the energy-dependent delta-shell pseudopotential follows as above, parameterized by some energy $E_0$,
\begin{equation}
V_{l,E_0}(r )
 = -\frac{g_{l,E_0}(r_s)}{f_{l,E_0}(r_s)} \beta_l(E_0) \delta(r-r_s)   \left[ -\frac{g'_{l,E_0}(r_s)}{g_{l,E_0}(r_s)}+\frac{\partial}{\partial r} \right] .
\label{potentialnewregularization}
\end{equation}
For the case of harmonic isotropic traps, with energy eigenvalues $E_0=2\nu_0+l +3/2$, the Kummer function radial eigenfunctions $F_{l,E_0}(r) = r^l e^{-r^2} w_{l,\nu_0}(r)$ defined by Eq.\ (\ref{Kummer}), can be put into the canonical form, Eq.\ (\ref{eigenfunctions}), by choosing 
\begin{subequations}
\begin{align}
f_{l,E_0}(r) &= r^l e^{-r^2} M(-\nu_0,l+3/2,r^2), \\
g_{l,E_0}(r) & = -r^{-(l+1)} e^{-r^2}  \frac{[(2l+1)!!]^2}{ 2l+1} \\
& M(-\nu_0-l-1/2,-l+1/2,r^2).\nonumber
\end{align}
\end{subequations}
Note that the pseudo potential depends on the trapping potential through the energy of the wave functions $E_0=2\nu_0+l+3/2$.  Thus, the wave functions at any other energies but the bound state energies in the trap will exponentially increase in the classically forbidden region. Since we evaluate the functions only at the shell radius $r_s$ in the construction of the pseudopotential, this is not a problem.  Moreover, our pseudopotential, by construction, only yields the correct asymptotic wave function at the energy $E_0$ which do not have to be equal to the eigenvalue in the trap.  To deal with this, we must set $E_0$ equal to $E$ self-consistently, as we discuss in the following section.

To calculate the biorthonormal set of wave functions, we need the adjoint of the potential. Employing $\braket{\Phi}{A\Psi}=\braket{A^\dagger\Phi}{\Psi}$, the adjoint potential is calculated to be
\begin{equation}
V_l^\dagger(z)=c_1 \tan{(\delta_l)}\left[c_2 \delta(r-r_s)-\frac{\partial}{\partial r}(\delta(r-r_s)\; )\right].
\label{adjointpotential}
\end{equation}
The boundary conditions for the adjoint set of wave functions, denoted $P(r)$, are obtained in a manner similar Eq. (\ref{boundaryfromschroedinger}), yielding
\begin{equation}
\frac{1}{2}\left.\left[\left.\frac{\partial}{\partial r} P^>\right|_{r_s} - \left. \frac{\partial}{\partial r} P ^< \right|_{r_s} \right]  + c_1 c_2 \tan{\delta_l(k)} P^>\right|_{r_s} =0.
\label{boundaryadjointderivative}
\end{equation}
Because the adjoint potential contains a derivative of a delta function, the adjoint wave functions are discontinuous at $r=r_s$, as can be seen by integrating once more,
\begin{equation}
\left.\left.\frac{1}{2}P^>\right|_{r_s}-\left.\frac{1}{2}P^<\right|_{r_s}-c_1\tan{\delta_l(k)}P^<\right|_{r_s}=0.
\label{boundaryadjointwavefunction}
\end{equation}
For $r>r_s$, however, these wave functions have the exact same behavior as $F(r)$.

As an example, consider two particles in an infinite spherical box of radius $R$, interacting through the potential 
$V=u\delta(r-r_s)\partial/\partial r$. Integrating the Schr\"{o}dinger equation over the potential yields the following boundary condition for the derivative of the wavefunction.
\begin{equation}
\left.\frac{\partial}{\partial r}\Psi^+\right|_{r_s}-\left.\frac{\partial}{\partial r}\Psi^-\right|_{r_s}+u\left.\frac{\partial}{\partial r}\Psi^+\right|_{r_s}=0,
\end{equation}
The adjoint potential is given by $\partial^-/\partial r(u \delta(r-r_s)\;)$, where $\partial^-/\partial r$ signifies the derivative acting to the left. This leads to a discontinuous adjoint wave function, with
\begin{equation}
-\left.\Phi^+\right|_{r_s}+\left.\Phi^-\right|_{r_s} +\left. u \Phi^-\right|_{r_s}=0.
\end{equation}
Both the wave function of the original potential and the derivative of the adjoint wavefunction are continuous, as can be seen in Fig.\ \ref{fig:deltainbox}. Outside the radius of the shell the wave functions are identical, while the inside wave functions differ by their normalisation.

\begin{figure}
\begin{center}
\includegraphics[width=9cm]{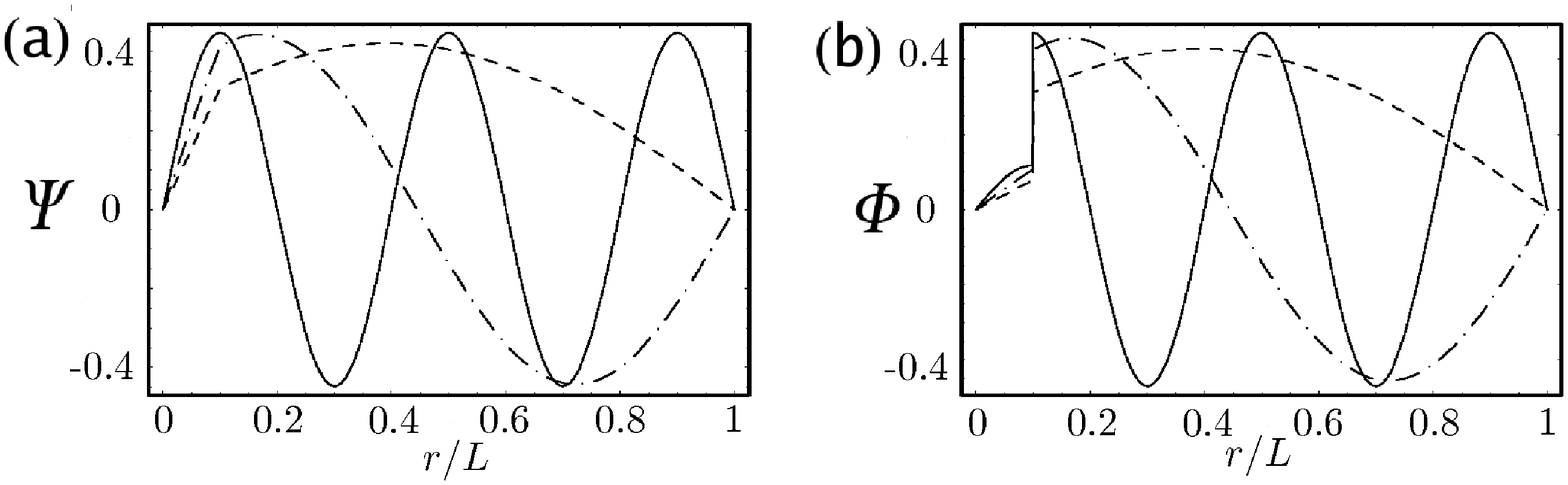}
\caption{(a): The first (solid), second (dashed) and fifth (dash-dotted) wave function of the potential $V=u\delta(r-r_s)\partial/\partial r$ in a one dimensional infinite box of arbitrary length $L$, with $r_s=0.1L$; (b) : the discontinuous wave functions of the adjoint potential, $V^\dagger=\partial^-/\partial r(u \delta(r-r_s) .)$. The wave functions form a complete biorthonormal set.}
\label{fig:deltainbox}
\end{center}
\end{figure}

The biorthogonal eigenfunctions obtained from Eqs (\ref{boundaryadjointderivative},\ref{boundaryadjointwavefunction}) can be specialized to the case of atoms in a harmonic trap. In terms of the regular and irregular functions they are given by
\begin{subequations}
\begin{align}
P_{l,E}^{<}(r) &=\frac{1/2 A_l(1+\beta_l(k) g_{l,E}(r_s)/f_{l,E}(r_s))}{1/2+c_1 \tan{(\delta_l)}} f_{l,E}(r),  \\
P_{l,E}^{>}(r) &=  A_l(f_{l,E}(r) +  \beta_l(k) g_{l,E}(r) ).
\label{biorthoeigenequations}
\end{align}
\end{subequations}
This can be obtained easily from the eigenfunctions of the potential, Eqs.\ (\ref{eigenfunctions}).

\section{Self-consistent solution for the quasi-Hermitian pseudopotential}
\label{sec:doublyselfconsistent}

Associated with the pseudopotential, Eq. (\ref{potentialnewregularization}),  is a biorthogonal set of eigenfunctions parameterized by an energy $E_0$.  We seek the eigenfunctions for energies $E$ that are the self-consistent solutions for two atoms interacting in the trap.  Naively, one may attempt to use the self-consistent eigenvalues found from the Busch solutions, Eq. (\ref{general_Busch}}), as the self-consistent parameter $E_0$.  These resulting wave functions do not, however, give rise to a biorthogonal set since the regularization will vary for each eigenfunction.  Instead, we must {\em fix} the parameter $E_0$ in Eq. (\ref{potentialnewregularization}) at an arbitrary value to find a given biorthogonal set, and then find the value of $E_0$ that allows us to satisfy all the boundary conditions in the trap.  A side effect of this procedure is that for energies $E \neq E_0$, the scattering phase shift arising from the pseudopotential is {\em not equal} to the scattering phase shift of the true potential at that energy.  We are thus faced with the task of ensuring self-consistent scattering phase shifts at the same time as we satisfy the trap's boundary condition.  We achieve this through a dual parameterization as discussed below.

\begin{figure}
\begin{center}
\includegraphics[width=8cm]{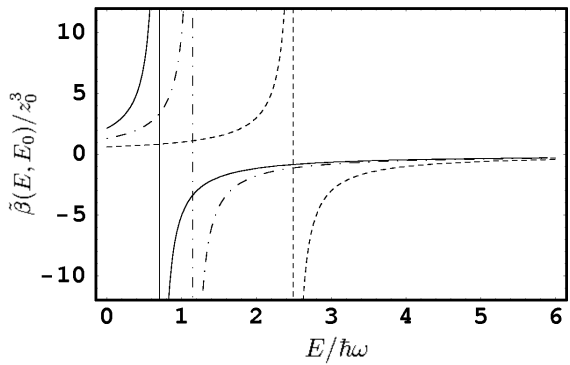}
\caption{The  dressed scattering length $\tilde{\beta}(E,E_0)$ that is imposed on the wave function by the pseudopotential as a function of the energy, for different bare scattering lengths $\beta(E)$ and $E_0=1$. Solid line: $\beta(E)=-5$, dashed: $\beta(E)=1$; dashed-dotted: $\beta(E)=10$. For $E = E_0$ the bare and dressed scattering lengths agree, $\tilde{\beta}(E,E_0)=\beta(E)$, and for other energies there is no agreement. For fixed $E_0$ the position of the pole depends on the bare scattering length of the interaction potential, while the width of the pole depends on the shell radius $r_s$.  All units are normalized to the characteristic scales of the harmonic oscillator as discussed in the text.}
\label{fig:betaofenergy}
\end{center}
\end{figure}

Consider the solutions at an energy $E$ of the pseudopotential delta-shell parameterized by an unknown energy $E_0$.  Given the solution outside the shell in canonical form, $A_l(f_l(E)-\tilde{\beta} g_l(E))$, and the boundary conditions, Eq. (\ref{boundaryfromschroedinger}), the scattering length function is,
\begin{multline}
\tilde{\beta}_l (E,E_0)=\\
\frac{(f_{l} \, g'_{l0} - f'_{l} \, g_{l0})\beta_l(E_0)}{(f_{l0}/f_{l}) (f_{l} \, g'_{l}-f'_{l}\, g_{l})+( g_{l0} \, g'_{l}- g_{l} \, g_{l0}')\beta_l(E_0) },
\label{beta}
\end{multline}
where the regular and irregular functions, $f_l$ and $g_l$ are evaluated at the shell radius, $r=r_s$. Here, $g_{l0}=g_{l,E_0}$, $g_l=g_{l,E}$, $f_{l,0}=f_{l,E_0}$, $f_l=f_{l,E}$, and $g'_{l0}=g'_{l,E_0}$ denotes the radial derivative.  The scattering phase shift that is imposed on the wave function by the pseudopotential is strongly energy dependent, matching the true interaction potential only when $E=E_0$, so that $\tilde{\beta}_l(E=E_0, E_0) =\beta_l(E_0)$. 
To better understand the behavior of the scattering length coefficient, consider the case of free-space collisions where $f_{l,E}=j_l(kr_s)$ and $g_{l,E}=n_l(kr_s)$.  In the limit as $r_s\rightarrow 0$, the $s$-wave and $p$-wave scattering length coefficients that follow from Eq.\ (\ref{beta}) are,
\begin{subequations}
\begin{align}
\tilde{\beta}_0(E,E_0) & =  \frac{k \beta_0(E_0) }{k_0+r_s  \left( k_0^2 - k^2 \right) \beta_0(E_0)}, \\
\tilde{\beta}_1(E,E_0) & = \frac{3 k^3 \beta_1(E_0) }{2 k_0^3 + k k_0^2 - \frac{3}{r_s} \left( k_0^2 - k^2 \right) \beta_1(E_0)}
\label{betatildeofbeta}.
\end{align}
\end{subequations}
Note the existence of a pole in $\tilde{\beta}$ at an energy depending on the shell radius.  In the case of $s$-waves, the position of the pole goes to $k=\infty$ as $r_s \rightarrow 0$ and we recover $\tilde{\beta}_0(E,E_0)  = (k/k_0) \beta_0(E_0)$ for finite energies.  Thus, for $s$-waves, one can construct a contact potential simply using the bare energy-dependent scattering length.  In contrast, for $p$-waves, the limit $r_s \rightarrow 0$ is singular.  The pole occurs at $k=k_0$ and the scattering length is indeterminate.  One requires, therefore, a pseudopotential for higher partial waves with finite radius $r_s$.  For the case of two atoms in a harmonic trap interacting via $p$-wave collisions, the outside wave function is given by the Kummer $U$ function. Eq.\ (\ref{beta}) then reduces for small $r_s$ to,
\begin{equation}
\tilde{\beta}_1(E,E_0) \rightarrow \frac{3 \beta_1(E_0)}{3+\frac{4}{r_s} (E_0-E)\beta_1(E_0) }.
\label{betaofgammasimplified}
\end{equation}
The  width of the pole depends inversely on the shell radius $r_s$. An example of $\tilde{\beta}_l(E,E_0)$ with fixed $E_0$ can be seen in Fig.\ref{fig:betaofenergy}.

We can now calculate a self-consistent solution for interacting atoms in the same trap.  For a given interaction potential we calculate (analytically where possible, numerically otherwise) the bare scattering length functions $\beta_l(E_0) = -\tan(\delta_l(E_0))/k_0^{2l+1}$.  Using  Eq.\ (\ref{beta}) we find the dressed scattering length function of the pseudopotential $\tilde{\beta}_l(E,E_0)$.  The generalized Busch eigenvalues are then found from Eq.\ (\ref{general_Busch}), with $\beta_l(E) \rightarrow \tilde{\beta}_l(E,E_0)$,
\begin{equation}
\tilde{\beta}_l(E,E_0)= (-1)^l \frac{2}{\pi} \left[ \frac{\Gamma(l+3/2)}{(2l+1)!!} \right]^2 \frac{\Gamma(-\nu-l-1/2)}{\Gamma(-\nu)}.
\label{final_Busch}
\end{equation}
The solutions of this transcendental equation with $E=2\nu+l+3/2$ can be found numerically for arbitrary $E_0$.  Given the energy eigenvalues, one can determine the biorthonormal basis, parameterized by $E_0$, using the Kummer equations and boundary conditions Eqs. (\ref{boundaryfromschroedinger},\ref{boundaryadjointwavefunction}).  For two atoms in the same trap, the self-consistent solution follows by setting $E_0$ equal to these eigenvalues.  In the next section, we solve the problem of two interacting  atoms in separated traps via this self-consistent procedure.

\section{Trap-induced resonances for higher partial wave scattering}
\label{sec:tisr}

We consider the application of our formalism to study the interaction of two atoms, trapped separately in displaced harmonic wells.  We assume each well is seen only by one of the distinguishable atoms, as in the state-dependent traps associated with a polarization gradient lattice~\cite{deutsch1998}.  The Hamiltonian of the system is
\begin{multline}
H=\frac{\mathbf{p}_1^2}{2m}+\frac{\mathbf{p}^2_2}{2m}+
\frac{1}{2}m\omega^2\left(\mathbf{r}_1+\frac{\Delta \v{z}}{2}\right)^2+\\
\frac{1}{2}m\omega^2\left(\mathbf{r}_2-\frac{\Delta \v{z}}{2}\right)^2
+V_\mathrm{int}(\mathbf{r}_1-\mathbf{r}_2),
\end{multline}
where $\Delta \v{z}$ is the vector separating the traps and $V_\mathrm{int}(\mathbf{r}_1-\mathbf{r}_2)$ is the true interaction potential between the atoms, which we will model using our pseudopotential. Due to the quadratic form of the Hamiltonian, one can separate the noninteracting center-of-mass from the relative coordinate.  With our choice of units, i.e. $\hbar=1$, $\mu=m/2=1$, $\omega=1$ and the characteristic length $z_0=\sqrt{\hbar/(\mu \omega)}$,
this is governed by the following Hamiltonian.
\begin{subequations}
\begin{align}
H_\mathrm{rel}=H_0 + H(\Delta z),  \\
H_0 \equiv \frac{1}{2}\left(-\frac{1}{r}\frac{\partial^2}{\partial r^2}r+\frac{l(l+1)}{r^2}+r^2\right)+\hat{v}_\mathrm{int}(r) ,\\
H(\Delta z) \equiv -\Delta zr\cos{\theta}+\frac{1}{2}\Delta z^2,
\end{align}
\label{hamiltonian}
\end{subequations}
where $\theta$ is the polar angle between the relative coordinate and $\Delta \v{z}$.  We have studied this problem previously in the case of $s$-wave collisions and observed the existence of trap-induced resonances when the potential energy of the trap shifts a molecular bound state of the interaction into resonance with a vibrational state of the trap \cite{stock2003}. Here, we study the generalization to higher partial wave scattering.

\begin{figure}
\begin{center}
\includegraphics[width=7cm]{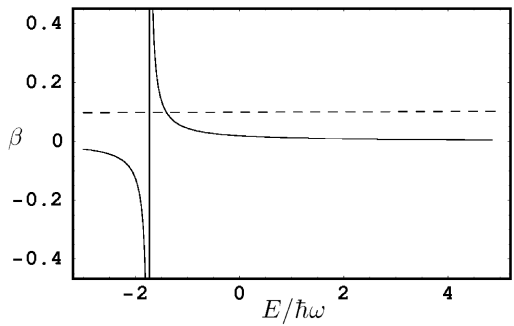}
\caption{Partial wave scattering lengths for the square well potential as a function of energy in harmonic oscillator units. Dashed line: $s$-wave scattering length (in units of the characteristic length of a harmonic oscillator, $z_0$);  Solid line: the scattering volume $\beta(E)$ for $p$-waves in units of $z_0^3$.}
\label{fig:a0_and_a1}
\end{center}
\end{figure}

A numerical solution follows from the quasi-Hermitian pseudopotential construction. We can construct a complete biorthogonal set from the eigenfunctions of $H_0$ with $\hat{v}_\mathrm{int}$ set to the delta-shell pseudopotential, parameterized by an unknown energy $E_0$ and scattering phase shift $\delta_l(E_0)$ according to the procedure in Sec.\ (\ref{sec:doublyselfconsistent}).  We use these functions to find a matrix representation of the perturbation $H(\Delta z)$.  Diagonalizing yields the total eigenvalue $E$.  The self-consistent solution is the one where $ E =E_0+ \Delta z^2 /2$, so that $E_0$ is the kinetic energy of the collision.

\begin{figure}
\begin{center}
\includegraphics[width=9cm]{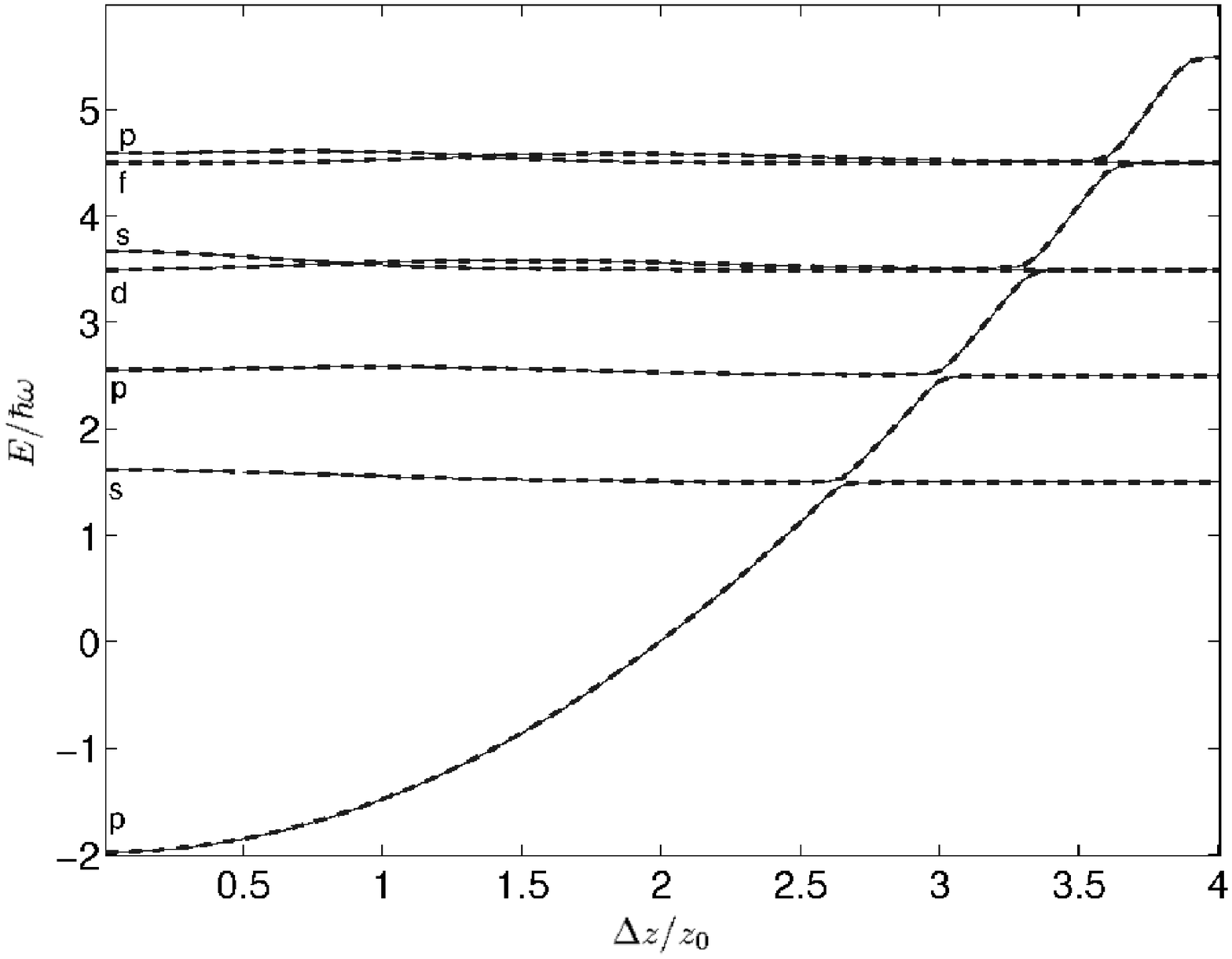}
\caption{Eigenspectrum for two atoms in displaced harmonic oscillators as a function of trap separation, for the example of a square-well interaction potential with a radius of $R_0=0.1$ and depth $V_0=489.9$ in harmonic oscillator units. Only the states with angular momentum projection in the direction of the separation $m=0$ are plotted. Interaction between higher partial waves than $p$-waves were neglected. The slash-dotted line is the solution based on the biorthonormal eigenfunctions of the pseudopotential, while the thin line shows the numerical solution to the exact problem. Each state is denoted by its angular momentum $l$ for no trap separation. For finite $\Delta z$ the states are mixed due to the separation. Several trap-induced resonances due to the shifted molecular $p$-wave boundstate are visible. The agreement is excellent.}
\label{fig:agreement}
\end{center}
\end{figure}

We test this procedure with a toy model of the molecular binding potential: a spherically symmetric step potential, of depth $V_0$ and radius $R_0$, whose scattering phase shift can be found analytically.  A plot of the scattering length can be seen in Fig.\ \ref{fig:a0_and_a1}. We compare the self-consistent solutions based on the pseudopotential with those generated from an orthonormal basis -- eigenstates of  $H_0$ with $v_\mathrm{int}$ taken to be the exact potential.  The latter basis is determined by matching Kummer function radial wave functions across the step radius as in \cite{stock2005}.  In both cases, we neglect scattering beyond $p$-wave. We only show states with angular momentum projection $m=0$ along the direction of the trap separation. Due to the axial symmetry of the problem, the states with different $m$ are not coupled to each other.

The agreement is excellent as can be seen in Fig.\ \ref{fig:agreement}.  The plot shows some interesting features of trapped separated atoms. For our step-well, there is a $p$-wave bound state at energy $E=-2$; the $s$-wave bound state is very deep and not shown. The $p$-wave bound state is shifted up quadratically by the trap, leading to trap-induced resonances where it crosses the higher lying vibrational states, due to the mixing of the partial waves at finite separation described below. Note that it only mixes with one of the states of each degenerate subspace, leading to both an avoided and an actual crossing for each of the degenerate manifolds.  At $\Delta z =0$, the higher lying states are $p,\;d,\;s,\;f$ and $p$ states, respectively, as noted in Fig. \ref{fig:agreement}.  Only $s$ and $p$ states are shifted in our approximation.  For finite separation, the vibrational states are superpositions of partial waves with respect to the interaction center.  In each subspace of vibrational states that has a set of degenerate partial waves with respect to the trap center without the interaction, there is a linear transformation of the degenerate eigenspace such that  the $j^{\text{th}}$ eigenvector contains no partial wave components for $l=0\dots j-2$. Thus, as seen in our previous work, \cite{stock2003}, with only $s$-wave interactions, the second excited vibrational manifold contained one state which is unshifted.  In our case, this state is affected by $p$-wave scattering at finite separation. In general, including up to $k$ partial waves in the interaction, $k$ states within a degenerate subspace will be shifted.  A small avoided crossing due to higher-order interactions can be seen between these states at a separation of $0.9$. Similar behavior is seen for the third vibrational subspace, with an avoided crossing between the sixth and seventh state at a separation of $\Delta z = 1.3$. Mixing of partial wave solution is generally seen in anisotropic traps \cite{bolda2003}.

\section{Summary}
\label{sec:summary}
In this paper, we considered a quasi-Hermitian pseudopotental for higher partial wave scattering based on a delta-shell potential with energy-dependent regularization.  In contrast to all other generalizations for $l>0$, the regularization derived here contains only a first order derivative, rendering it quasi-Hermitian and supporting a complete biorthogonal set of eigenfunctions.   Such a set is useful for studying interactions of atoms in tightly confining traps as it forms a basis for easily diagonalizing the relevant Hamiltonian. Because the regularization is energy dependent, we must solve such problems self-consistently, so that the eigenvalues of the diagonalization match those of the parameters in the pseudopotential.

We have tested our construction by studying the problem of two atoms trapped in separated wells, with a square well to model the inter-atomic interaction.  The solutions obtained from our pseudopotential show excellent agreement with the numerical solutions to the complete problem. We also showed that a $p$-wave bound state close to dissociation leads to trap-induced resonances, which have been calculated earlier for $s$-wave scattering. This is particularly interesting for identical fermions which cannot interact via $s$-wave collisions due to their quantum statistics.

In future research, we plan to apply our method to study realistic atomic systems.  A particular relevant case is cesium collisions.  Due to a bound state of the cesium dimer almost exactly at dissociation, the collision properties are anomalous.  We have already seen large trap-induced resonances for $s$-wave scattering in previous theoretical studies \cite{stock2006}.  Because cesium has large spin-relaxation, high partial wave scattering can play an important role.  Trap-induced resonances should then be exhibited in multichannel scattering processes.

\appendix*
\section{Hermiticity of the regularized pseudopotential for $s$-waves}
\label{$s$-wavehermiticity}
The regularized Fermi pseudopotential for $s$-waves, given by Eq.\ (\ref{swavepseudopotential}), is not a Hermitian operator.  Nor is the kinetic energy operator when its domain is extended to include the space of the irregular functions at the origin \cite{merzbacher}. Surprisingly the non-Hermitian parts of these operators cancel exactly, as can be seen as follows. The antihermitian part of our Hamiltonian is
\begin{multline}
H_{A_{i,j}}=\int^{\infty}_0(F^*_iHF_j-F_jHF^*_i)\: dr=\\
\frac{1}{2}\left[ rF_j\frac{\partial}{\partial r}(r F_i^*)-a_i F_j\frac{\partial}{\partial r}(r F_i^*)- \right.\\
\left.\left.(r F_i^*)\frac{\partial}{\partial r}(r F_j)+a_jF_i^*\frac{\partial}{\partial r} (rF_j)\right]\right|_0
\end{multline}
Here, $a_i$ and $a_j$ are the energy-dependent scattering lengths at the eigenenergies $2 \nu_i+3/2$ and $2 \nu_j +3/2$ for $F_i$ and $F_j$, respectively. The wave functions are Kummer $U$-functions, given by $C e^{-r^2}(M(-\nu_i,3/2,r^2)-a_i 1/r M(-\nu_i-1/2,1/2,r^2))$, where $C$ is a constant. Pulling out the factor of $e^{-r^2}$ and using the fact that $M(a,b,r)=1+ar/b+...,$ we obtain
\begin{equation}
H_{A_{i,j}}-a_j+a_i-a_i+\frac{a_ia_j}{r}+a_j-a_i-\frac{a_ia_j}{r}=0.
\end{equation}
Since this is zero for all $i,j$, the Hamiltonian for the $s$-wave pseudopotential is Hermitian. Thus the eigenfunctions form an orthonormal set. No such cancellation can occur for higher partial waves, as discussed in Sec. \ref{sec:non-Hermitian}


\begin{acknowledgments}
We thank Satyan Bhongale for helpful discussions. This work was partly supported by ARDA Contract No.~DAAD19-01-1-0648 and by the ONR Contract No.~N00014-03-1-0508. R. S. also acknowledges support by the Alberta’s Informatics Circle of Research Excellence (iCORE).
\end{acknowledgments}

\bibliography{pseudopotential}

\end{document}